\documentclass[conference]{IEEEtran}
\IEEEoverridecommandlockouts
\usepackage{comment}
\usepackage{mathtools}
\usepackage{bbm}
\usepackage{graphicx}
\usepackage{lipsum}
\usepackage{rotating}
\usepackage{epsfig,graphics,amssymb}
\usepackage{amscd}
\usepackage{amsmath}
\usepackage{enumerate}
\usepackage{theorem}
\usepackage{cite}
\usepackage{stfloats}
\usepackage{epsfig}
\usepackage{verbatim}
\usepackage{color}
\theoremstyle{plain} \theorembodyfont{\itshape}
\theoremheaderfont{\scshape}

\theorembodyfont{\itshape} \theoremheaderfont{\scshape}

\theoremstyle{plain} \theorembodyfont{\itshape}
\theoremheaderfont{\scshape}

\newtheorem{assumption}{Assumption}

\newtheorem{remark}{Remark}

\title{S-AMP for Non-linear Observation Models}
\author{
\IEEEauthorblockA{Burak \c{C}akmak \\ Department of Electronic Systems \\ Aalborg University \\ 9220 Aalborg, Denmark\\Email: buc@es.aau.dk} \and  
\IEEEauthorblockA{Ole Winther\\ DTU Compute \\Technical University of Denmark\\2800 Lyngby, Denmark \\Email: olwi@dtu.dk}  \and 
\IEEEauthorblockA{Bernard H. Fleury \\ Department of Electronic Systems \\ Aalborg University \\ 9220 Aalborg, Denmark\\Email: fleury@es.aau.dk} }
\begin{document}
\def\mathlette#1#2{{\mathchoice{\mbox{#1$\displaystyle #2$}}%
                               {\mbox{#1$\textstyle #2$}}%
                               {\mbox{#1$\scriptstyle #2$}}%
                               {\mbox{#1$\scriptscriptstyle #2$}}}}
\newcommand{\matr}[1]{\mathlette{\boldmath}{#1}}
\newcommand{\RR}{\mathbb{R}}
\newcommand{\CC}{\mathbb{C}}
\newcommand{\NN}{\mathbb{N}}
\newcommand{\ZZ}{\mathbb{Z}}
\newcommand{\LL}{\rotatebox[origin=c]{180}{$\Lambda$}}
\newcommand{\bfl}[1]{{\color{blue}#1}}
\maketitle
\begin{abstract}
Recently we extended Approximate message passing (AMP) algorithm to be able to handle general invariant matrix ensembles. In this contribution we extend our S-AMP approach to non-linear observation models. We obtain generalized AMP (GAMP) algorithm as the special case when the measurement matrix has zero-mean iid Gaussian entries. Our derivation is based upon 1) deriving expectation propagation (EP) like algorithms from the stationary-points equations of the Gibbs free energy under first- and second-moment constraints and 2) applying additive free convolution in free probability theory to get low-complexity updates for the second moment quantities.
\end{abstract}
\section{Introduction}
Approximate message passing techniques, e.g. \cite{Donoha1,Florent,Rangan}, have recently received significant attention by the signal processing community. Essentially, these methods are based on taking the large system limit of loopy belief propagation where the central limit theorem can be applied when the underlying measurement matrix has independent and zero-mean entries.

Variational inference techniques are well-established in the field of information theory e.g. \cite{Yedida}, \cite{Erwin} and machine learning e.g. \cite{Heskes,Ole5}. For example, it is well-known that exact inference can be formulated as the solution to a minimization problem of the Gibbs free energy of the underlying probabilistic model under certain marginalization consistency constraints \cite{Yedida}. We have recently shown in \cite{Samp} that for the zero-mean independent identically distributed (iid) measurement matrix, approximate message passing (AMP) algorithm \cite{Donoha1} can also be obtained from the stationary-points equations of the Gibbs energy under first- and second-moment consistency constraints. Furthermore, AMP can be extended to general \emph{invariant}\footnote{Note that we omit to mention the invariance property in \cite{Samp}. It is however crucial for the derivation.} matrix ensembles by means of the asymptotic spectrum of the measurement matrix. We call this approach S-AMP (where S comes from the fact that the derivation uses the S-transform).

AMP is an estimation algorithm for the linear observation models. However many interesting cases occur in practice where the observation model is non-linear, e.g non-linear form of compressed sensing, Gaussian processes for classification. In this article we extend S-AMP approach \cite{Samp} to general observation models. Specifically we address the sum-product generalized AMP (GAMP for short) algorithm\cite{Rangan}.

The derivation of GAMP is based on certain approximations (mainly Gaussian and quadratic approximations) of loopy belief propagation. If the measurement matrix is large and has zero mean and iid entries, GAMP provides excellent performance, e.g. \cite{Rangan,1bit}. Furthermore, for general matrix ensembles it can show quite reasonable accuracy \cite{swept}. However the algorithm itself and its derivation are not well-understood.

To better understand GAMP, in \cite{Rangan2} the authors characterize its fixed points. Specifically, they show that GAMP can be obtained from the stationary-point equations of some implicit approximations of naive mean-field approximation \cite{Rangan2}. These implicit approximations only provide limited insight. Furthermore, the naive mean-field interpretation is misleading, because the fixed points of AMP-type algorithms are typically known as the TAP-like equations, i.e. they include a correction term to naive mean-field solution. In fact GAMP can also be obtained from the stationary-points equations of the Bethe free energy (BFE) of the underlying loopy graph under first- and second-moment constraints. However, this approach also limits our understanding, because the BFE formulation of a loopy graph is suitable for sparsely connected systems.

In this work we focus on the BFE formulation of a tree graph, i.e. an exact Gibbs free energy formulation. We note that our approach coincides with the expectation prorogation (EP) \cite{Ole0,Minka1,Ole} since the fixed points of EP are the stationary points of BFE of the underlying probabilistic graph under a set of moment consistency constraints\cite{Heskes}.

\subsubsection*{Notations} The entries of the $N\times K$ matrix $\matr X$ are denoted by either $X_{nk}$ or $[X]_{nk}$, $n\in \mathcal N\triangleq \{n:1\leq n\leq N\}$ and $k\in \mathcal K\triangleq\{k: 1\leq k\leq K\}$. Without loss of generality we assume that $\mathcal K$ and $\mathcal N$ are disjoint. $(\cdot)^\dagger$ denotes the transposition. We denote by $\Re z$ and $\Im z$ the real and imaginary part of $z\in \CC$, respectively. The entries of a vector $\matr u\in \mathbb R^{T\times 1}$ are indicated by either $u_t$ or $[u]_t$, $t \in [1, T]$. Furthermore $\left<\matr u\right>\triangleq\sum_{t=1}^{T} u_t/T$. Moreover, ${\rm diag}(\matr u)$ is a diagonal matrix with the elements of vector $\matr u$ on the main diagonal. For a square matrix $\matr X$, ${\rm diag}(\matr X)$ is a column vector containing the diagonal elements of $\matr X$. Furthermore ${\rm Diag}(\matr X)\triangleq {\rm diag}({\rm diag}(\matr X))$. The Gaussian probability density function (pdf) with mean $\matr \mu$ and the covariance $\matr \Sigma$ is denoted by $N(\cdot ; \matr \mu, \matr\Sigma)$. Throughout the paper when referring to ``in the large system limit'' we imply that $N,K$ tends to infinity with the ratio $\alpha\triangleq N/K$ fixed. All large system limits are assumed to hold in the almost sure sense, unless explicitly stated.

\section{System Model and Review of GAMP}
Consider the estimation of a random vector $\matr x \in \RR^{K\times 1}$ which is linearly transformed by $\matr A \in \RR ^{N\times K}$ as $\matr z\triangleq \matr A\matr x$, 
then passed through a noisy channel whose output is given by $\matr y\in \RR^{N\times 1}$. We assume that the conditional pdf of the channel factorizes according to 
\begin{equation}
p(\matr y\vert \matr z)=\prod_{n\in \mathcal N}p(y_n\vert z_n).
\end{equation}
Furthermore for the Bayesian setting we assign a prior pdf for $\matr x$ that is assumed to be factorized as 
\begin{equation}
p(\matr x)= \prod_{k\in \mathcal K}p(x_k).
\end{equation}

\subsection{\rm GAMP summarized}
We summarize GAMP here for the sake of streamlining and making the connection to the derivation of S-AMP. We separate the GAMP iteration rules \cite{Rangan} into two parts: (i) GAMP-1st order that initializes $\matr {\hat x}^{t}$, $\matr \tau_{\rm x}^{t}$ and $\matr m^{t}$ from tabula rasa at $t\leq 0$ and proceeds iteratively as
\begin{align}
\matr \kappa_{\rm z}^t&=\matr A\matr{\hat x}^t-(\matr\LL_{\rm z}^{t})^{-1}\matr m^{t-1} \label{G1}\\
\matr {\hat z}^t&=\mu_{\rm z}(\matr \kappa_{\rm z}^t;\matr \LL_{\rm z}^t)  \label{G2} \\
\matr \tau_{\rm z}^t&= \sigma_{\rm z}(\matr \kappa_{\rm z}^t;\matr\LL_{\rm z}^t) \label{G3} \\
\matr m^t&= \matr\LL_{\rm z}^t (\matr{\hat z}^t-\matr\kappa_{\rm z}^t)\label{G4} \\
\matr\kappa_{\rm x}^t&=(\matr\LL_{\rm x}^t)^{-1}\matr A^\dagger\matr m^t+\matr{\hat x}^t \label{G5}\\
\matr {\hat x}^{t+1}&=\mu_{\rm x}( \matr\kappa_{\rm x}^t;\matr\LL_{\rm x}^t) \label{G6} \\
\matr \tau_{\rm x}^{t+1}&=\sigma_{\rm x}(\matr\kappa_{\rm x}^t;\matr\LL_{\rm x}^t)\label{G7}.
\end{align}
(ii) GAMP-2nd order are the update rules for $\matr\LL_{\rm z}^t$ and $\matr\LL_{\rm x}^t$:
\begin{align}
\matr\LL_{\rm z}^t&=({\rm diag}((\matr A \circ \matr A) \matr\tau_{\rm x}^t))^{-1}  \label{V1} \\
\matr \tau_{\rm m}^t&=\matr\LL_{\rm z}^t(\matr 1-\matr\LL_{\rm z}^t\matr \tau_{\rm z}^t)\label{V2} \\
\matr\LL_{\rm x}^t&={\rm diag}((\matr A \circ \matr A)^\dagger \matr \tau_{\rm m}^t).  \label{V3}
\end{align}
In these expressions $\matr 1$ is the all-ones vector of appropriate dimension and $\mu_{\rm x}$ and $\sigma_{\rm x}$ are scalar functions. Specifically, if $\matr\LL$ is a $K\times K$ diagonal matrix and $\matr \kappa$ is a $K\times 1$ vector; then for $k\in \mathcal K$, $[\mu_{\rm x}(\matr\kappa;\matr\LL)]_{k}$ and $[\sigma_{\rm x}(\matr\kappa; \matr \LL)]_k$ are respectively the mean and the variance taken over the pdf 
\begin{equation}
q_{k}(x_k)\propto p_k(x_k)\exp\left(-\frac{\LL_{kk}}{2}(x_k-\kappa_{k})^2\right) \label{qx}.
\end{equation}
Similarly, $\mu_{\rm z}$ and $\sigma_{\rm z}$ are scalar functions such that if $\matr\LL$ is a $N\times N$ diagonal matrix and $\matr \kappa$ is a $N\times 1$ vector, for $n\in \mathcal N$, $[\mu_{\rm z}(\matr\kappa;\matr\LL)]_{n}$ and $[\sigma_{\rm z}(\matr\kappa; \matr \LL)]_n$ are respectively the mean and the variance taken over the pdf
\begin{equation}
q_{n}(z_n)\propto p(y_n\vert z_n)\exp\left(-\frac{\LL_{nn}}{2}(z_n-\kappa_{n})^2\right)\label{qz}.
\end{equation}
If the entries of $\matr A$ are iid with zero mean and variance $1/N$, the iteration steps for the GAMP-2nd order simplify as
\begin{align}
\matr\LL_{\rm z}^t= \frac{\alpha}{\left<\matr\tau_{\rm x}^t\right>}{\bf I}, \quad \matr\LL_{\rm x}^t= {\left<\matr\tau_{\rm m}^t\right>}{\bf I}, \label{iid2}
\end{align}
where $\bf I$ is the identity matrix of appropriate dimension. We note that if in addition $p(\matr y\vert \matr z)=N(\matr y; \matr z,\sigma^2\matr {\bf I})$,  GAMP yields AMP, see e.g. \cite[Appendix~C]{Florent}.
\section{Gibbs Free Energy with Moment Constraints}
For the sake of notational compactness, consider $\matr s=(\matr x, \matr z)$. Furthermore we introduce the set $\mathcal V\triangleq \mathcal K \cup \mathcal N$ and assume that $\mathcal K$ and $\mathcal N$ are disjoint.
Moreover we define
\begin{align}
f_{A}(\matr s)&\triangleq\delta(\matr z-\matr A\matr x) \\
f_{v}(s_v)&\triangleq {\begin{cases}
p_v(x_v) & \: v \in \mathcal K\\
p(y_v\vert z_v) & \: v \in \mathcal N.
\end{cases}}
\end{align}
With these definitions, the posterior pdf of $\matr s$ reads
\begin{equation}
p(\matr s \vert\matr y, \matr A)=\frac{1}{Z}f_{A}(\matr s)\prod_{v \in \mathcal V}f_{v}(s_v)\label{fac}.
\end{equation}
with $Z$ denoting a normalization constant. The factor graph representing \eqref{fac} is a tree. Thus the BFE of \eqref{fac} is equal to its Gibbs free energy \cite{Yedida}:
\begin{align}
&{\rm G}(\{\tilde b_{v}, b_{A},b_{v}\})\triangleq-\sum_{v\in \mathcal V}\int \tilde b_{v}(s_v)\log \tilde b_{v}(s_v){\rm d}s_v \nonumber \\
&-\int b_{A}(\matr s) \log\frac{f_{A}(\matr s)}{b_{A}(\matr s)}{\rm d}\matr s -\sum_{v \in \mathcal {V}}\int b_{v}(s_v)\log\frac{f_v(s_v)}{b_{v}(s_v)}{\rm d}s_v\label{energy}. 
\end{align}
Here $b_{A}$ and $b_{v}$, $v\in \mathcal {V}$, denote the beliefs of the factors in \eqref{fac}, while $\tilde b_{v}$, $v\in \mathcal {V}$, denote the beliefs of the unknown variables in (\ref{fac}). Without loss of generality we assume that the expressions $f_{A}(\matr s)/b_{A}(\matr s)$ and $f_v(s_v)/b_{v}(s_v)$ in \eqref{energy} are strictly continuous; so that the Gibbs free energy is well-defined. Indeed this is what we will end up with in the analysis.    

If we define a Lagrangian for \eqref{energy} that accounts for certain marginalization consistency constrains, then at its stationary point, the belief $\tilde b_{v}(s_v)$ is equal to $p(s_v\vert \matr y,\matr A)$ for all $v \in \mathcal {V}$ \cite{Yedida}. Instead, following the arguments of \cite{Heskes}, we define the Lagrangian on the basis of a set of moment consistency constraints as
\begin{align}
&\mathcal L(\{\tilde b_{v}, b_{A},b_{v}\})\triangleq{\rm G}(\{b_{v}, b_{A},\tilde b_{v}\})+ \mathcal Z \nonumber \\
&-\sum_{v\in \mathcal V}\matr{\nu}_v^\dagger\int \matr \phi(s_v)\left\{b_A(\matr s)-\tilde b_v(s_v)\right\}{\rm d}\matr s\nonumber \\
&-\sum_{v\in \mathcal V}\matr {\bar \nu}_{v}^\dagger\int \matr \phi(s_v)\left\{b_{v}(s_v)-\tilde b_v(s_v)\right\}{\rm d}s_v.
\label{lagrage}
\end{align}
Here we consider constraints on the mean and variance, i.e. $\matr \phi(s_v)=(s_v,s_v^2)$, $v\in \mathcal {V}$. For convenience we write the Lagrangian multipliers as 
\begin{equation}
\matr{\nu}_v\triangleq\left(\gamma_v, -\frac{\Lambda_{vv}}{2} \right), \quad \matr{\bar \nu}_v\triangleq\left(\rho_v, -\frac{\LL_{vv}}{2} \right), 
\quad v\in \mathcal {V}\nonumber.
\end{equation}
The term $\mathcal Z$ accounts for the normalization constraints:
\begin{align}
&\mathcal Z\triangleq-\beta_A\left(1-\int b_{A}(\matr s){\rm d}\matr s\right) \nonumber \\
&-\sum_{v\in \mathcal V} \tilde \beta_v \left(1-\int \tilde b_v(x_v){\rm d}s_v\right)-\beta_v\left(1-\int b_{v}(s_v){\rm d}s_v\right)\nonumber
\end{align}
where $\beta_A$, $\beta_v$, $\tilde\beta_v$ are the associated Lagrange multipliers. 

We formulate the estimate of $s_v, v\in \mathcal V$, as 
\begin{equation}
\hat s_v\triangleq\int s_v \tilde b_v^{\star}(s_v)ds_v, 
\label{mmse}
\end{equation}
where $\tilde b_v^{\star}(s_v)$ represents $\tilde b_v( s_v)$ at a stationary point of \eqref{lagrage}.
\subsection{\rm The Stationary Points of the Lagrangian}\label{fixep}
For notational convenience we introduce first the $(K+N)\times (K+N)$ diagonal matrices $\matr{\Lambda}$  and $\matr {\LL}$ as well as the $(K+N)\times 1$ vectors $\matr{\gamma}$ and $\matr \rho$ whose entries are respectively $\Lambda_{vv}$, $\LL_{vv}$, $\gamma_v$ and $\rho_v$, $v\in \mathcal V$. In connection with variables $\matr x$ and $\matr z$ we write 
\begin{align}
&\matr \Lambda= \left( \begin{array}{cc}
\matr \Lambda_{\rm x}  & \matr 0 \\
\matr 0 & \matr \Lambda_{\rm z}
\end{array}\right), \quad \matr \gamma=(\matr \gamma_{\rm x}, \matr \gamma_{\rm z})\\
&\matr \LL= \left( \begin{array}{cc}
\matr \LL_{\rm x}  & \matr 0 \\
\matr 0 & \matr \LL_{\rm z}
\end{array}\right), \quad \matr \rho=(\matr \rho_{\rm x}, \matr \rho_{\rm z}). 
\end{align}
The dimensions of $\matr \Lambda_{\rm x}$ and $\matr \LL_{\rm x}$ are $K\times K$; vectors $\matr \gamma_{\rm x}$ and $\matr \rho_{\rm x}$ have dimension $K\times 1$.

Following the arguments of \cite{Heskes}, we have the stationary points of the Lagrangian \eqref{lagrage} in the form 
\begin{align}
\tilde b_{v}^{\star}(s_v)&=\frac{1}{\tilde Z_v}\exp ((\matr\nu_v+\matr{\bar\nu}_v)^\dagger\matr\phi (s_v)), \quad v\in \mathcal V\label{belief1}\\
b_{v}^{\star}(s_v)&= \frac{1}{Z_v}f_v(s_v)\exp (\matr{\bar\nu}_v^\dagger\matr\phi (s_v)),\quad v\in \mathcal V\label{belief}\\
b_{A}^{\star}(\matr s)&= \frac{1}{Z_{A}}f_{A}(\matr s)\exp\left(-\frac{1}{2}{\matr s^\dagger \matr \Lambda\matr s}+\matr s^\dagger \matr \gamma\right) \label{fullbelief}
\end{align}
where $Z_A$, $Z_v$, $\tilde Z_v$ are the associated normalization constants.

Let us first consider the marginalization of the belief $b_{A}^{\star}(\matr s)$ with respect to $\matr z$:
\begin{align}
b^{\star}_{A}(\matr x)&=\int b_{A}^{\star}(\matr x, \matr z) {\rm d}\matr z=N(\matr x; \matr {\hat x}, \matr \Sigma_{\rm x})
\end{align}
where 
\begin{equation}
\matr \Sigma_{\rm x}\triangleq (\matr \Lambda_{\rm x}+ \matr A^\dagger \matr \Lambda_{\rm z}\matr A)^{-1}, \quad \matr{\hat x}\triangleq\matr \Sigma_{\rm x}(\matr \gamma_{\rm x}+ \matr A^\dagger\matr \gamma_{\rm z}). \label{TAP1}
\end{equation}
Here we note that $\matr \Sigma_{\rm x}$ is positive definite since $b^{\star}_{A}(\matr x)$ is a well-defined pdf. Second, let us consider the marginalization over $\matr x$, which basically follows from the linear transformation property of a Gaussian random vector: 
\begin{align}
b_{A}^{\star}(\matr z)&=\frac{e^{-\frac{1}{2}\matr z^\dagger \matr{\Lambda}_{\rm z} \matr z+\matr z^\dagger \matr {\gamma}_{\rm z}}}{Z_A}\int\delta(\matr z-\matr A\matr x)e^{-\frac{1}{2}\matr x^\dagger \matr{\Lambda}_{\rm x} \matr x+\matr x^\dagger \matr {\gamma}_{\rm x}}{\rm d}\matr x \nonumber \\ 
&= \int\delta(\matr z-\matr A\matr x)N(\matr x;\matr {\hat x}, \matr \Sigma){\rm d}\matr x=N(\matr z;\matr {\hat z},\matr \Sigma_{\rm z})
\end{align}
where $\matr {\hat z}\triangleq\matr A\matr{\hat x}$ and $\matr \Sigma_{\rm z}\triangleq\matr A \matr \Sigma_{\rm x} \matr A^\dagger$.  

At this stage it is convenient to define
\begin{equation}
\matr\kappa \triangleq(\matr\kappa_{\rm x}, \matr \kappa_{\rm z})= (\matr \LL_{\rm x}^{-1} \matr\rho_{\rm x},  \matr \LL_{\rm z}^{-1} \matr\rho_{\rm z})
\end{equation}
with $\matr\kappa_{\rm x}\in \RR^K$. In this way we can write the belief in \eqref{belief} as
\begin{equation}
b_{v}^{\star}(s_v)\propto f_v(s_v)\exp\left(-\frac{\LL_{vv}}{2}(s_v-\kappa_v)^2\right)\label{kappab}.
\end{equation}
Thereby \eqref{kappab} has a form identical to \eqref{qx} and \eqref{qz} for $v \in \mathcal N$ and for $v\in \mathcal K$, respectively. Then let us define 
\begin{align}
\mu(\matr\kappa;\matr\LL)&\triangleq \left(\mu_{\rm x}(\matr\kappa_{\rm x};\matr\LL_{\rm x}),\mu_{\rm z}(\matr\kappa_{\rm z};\matr\LL_{\rm z}) \right) \\
\sigma(\matr\kappa;\matr\LL)&\triangleq \left(\sigma_{\rm x}(\matr\kappa_{\rm x};\matr\LL_{\rm x}),\sigma_{\rm z}(\matr\kappa_{\rm z};\matr\LL_{\rm z}) \right).
\end{align}
The entries $[\mu(\matr\kappa;\matr\LL)]_v$ and $[\sigma(\matr\kappa;\matr\LL)]_v$ are respectively the mean and variance of the belief \eqref{belief}. Moreover we introduce 
\begin{equation}
\matr \Sigma\triangleq\left( \begin{array}{cc}
\matr \Sigma_{\rm x}  & \matr 0 \\
\matr 0 & \matr \Sigma_{\rm z}
\end{array}\right), \quad \matr{\hat s}=(\matr {\hat x}, \matr {\hat z}).
\end{equation}
With these definitions, the identities resulting from the moment consistency constraints are given by
\begin{align}
\matr{\hat s}={\rm Diag}(\matr\Sigma)(\matr\gamma+\matr\rho), & \quad  \matr{\hat s}=\mu(\matr\kappa;\matr\LL) \label{f1}\\
{\rm Diag}(\matr\Sigma)=(\matr\Lambda+\matr\LL)^{-1}, & \quad  {\rm diag}(\matr\Sigma)=\sigma(\matr\kappa;\matr\LL).\label{f2} 
\end{align}
\subsection{\rm The TAP-like Equations and GAMP-1st Order}
By using the fixed-point identities presented in Section~\ref{fixep}, one can introduce numerous fixed-point algorithms. In this work we restrict our attention to TAP-like algorithms, e.g. \cite{Ole0, Ole, Rangan}. To that end we start with the definitions in \eqref{TAP1} and write
\begin{equation}
\matr \gamma_{\rm x}=-\matr A^\dagger \matr \gamma_{\rm z}+(\matr \Lambda_{\rm x}+ \matr A^\dagger \matr \Lambda_{\rm z}\matr A)\matr{\hat x}.
\end{equation}
Furthermore by making use of the identities in \eqref{f1} and \eqref{f2} we have 
\begin{align}
\matr \rho_{\rm x}&= \matr A^\dagger \matr \gamma_{\rm z}-(\matr \Lambda_{\rm x}+ \matr A^\dagger \matr \Lambda_{\rm z}\matr A)\matr{\hat x}+ (\matr \Lambda_{\rm x}+\matr \LL_{\rm x})\matr {\hat x} \\
&=\matr A^\dagger (\matr \gamma_{\rm z}- \matr \Lambda_{\rm z}\matr A\matr{\hat x})+ \matr \LL_{\rm x}\matr{\hat x}\\
&=\matr A^\dagger\matr m+\matr \LL_{\rm x}\matr{\hat x} \quad \mathrm{with}\quad \matr m\triangleq(\matr \gamma_{\rm z}- \matr \Lambda_{\rm z}\matr A\matr{\hat x}).
\end{align}
Moreover, by the definition of $\matr m$ we also point out that
\begin{align}
\matr m&=(\matr \Lambda_{\rm z}+\matr \LL_{\rm z})\matr{\hat z}-\matr{\rho}_{\rm z}- \matr \Lambda_{\rm z}\matr A\matr{\hat x} \\
&=\matr\LL_{\rm z}\matr {\hat z}-\matr \rho_{\rm z}=\matr\LL_{\rm z}(\matr {\hat z}-\matr \kappa_{\rm z}).
\end{align}
Thereby we exactly obtain the fixed points of GAMP-1st order, i.e. \eqref{G1}-\eqref{G7}. Now let us keep the iterations step of GAMP-1st order but define the update rule for $\matr\Lambda_{\rm x}^t$ and $\matr\Lambda_{\rm z}^t$ on the basis of the fixed point identities in \eqref{f2}. For example:
\begin{align}
\matr\Lambda_{\rm z}^t&= ({\rm diag}(\matr\tau_{\rm z}^{t-1}))^{-1}- \matr\LL_{\rm z}^{t-1}\label{EP1}\\
\matr\Sigma_{\rm x}^t&=(\matr \Lambda_{\rm x}^{t-1}-\matr A^\dagger \matr \Lambda_{\rm z}^t\matr A)^{-1}.\label{challenge}\\ 
\matr \LL_{\rm z}^t&= {\rm Diag} \left(\matr A \matr \Sigma_{\rm x}^t\matr A^\dagger\right)^{-1}- \matr \Lambda_{\rm z}^t\\
\matr\Lambda_{\rm x}^t&= {\rm diag}(\matr \tau_{\rm x}^{t})^{-1}- \matr\LL_{\rm x}^{t-1} \label{EP7}\\
\matr \LL_{\rm x}^t&= {\rm Diag} \left(\matr \Sigma_{\rm x}^t\right)^{-1}- \matr \Lambda_{\rm x}^t\label{EP8}.
\end{align}
In this way we obtain a new fixed point algorithm whose fixed points are the stationary point of Lagrangian \eqref{lagrage}. However from the complexity point of view these updates are problematic due to the matrix inversion in \eqref{challenge}. In the sequel we will address how to bypass \eqref{challenge} as $K,N$ are large. 
\subsection{\rm The Large-System Simplifications} \label{SectionLarge}
To circumvent the complexity problem \eqref{challenge}, we utilize the so-called additive free convolution in free probability theory \cite{hiai}. The reduction that we obtain in this way can be also obtained by means of the self-averaging ansatz in \cite[Section~3.1]{Ole}.

In order to make use of additive free convolution we need to restrict our consideration to the invariant matrix ensembles:\vspace{-0.2cm}
\begin{assumption}
Consider the singular value decomposition $\matr A= \matr U \matr D  \matr V$ where $\matr U^{N\times N}$ and $\matr V^{K\times K}$ are orthogonal matrices and $\matr D$ is a $N\times K$ non-negative diagonal matrix. We distinguish between the invariance assumption on $\matr A$ from right and from left: {\rm a)} $\matr A$ is invariant from right, i.e. $\matr V$ is Haar distributed; {\rm b)} $\matr A$ is invariant from left, i.e. $\matr U$ is Haar distributed.
\end{assumption}
It indeed makes sense to distinguish between the invariance from right and the invariance from left. For example, once we consider the classical linear observation model such as $p(\matr y\vert \matr z)=N(\matr y; \matr z,\sigma^2\matr {\bf I})$, then $\matr \Lambda_{\rm z}= \matr {\bf I}/\sigma^2$. In this case we do not need to consider Assumption~1-b). 

Second, we make the following technical assumption on the limiting spectrum of the respective matrices: \vspace{-0.1cm}
\begin{assumption}
As $N,K\to \infty$ with the ratio $\alpha=N/K$ fixed let the spectra of $\matr\Lambda_{\rm x}$, $\matr\Lambda_{\rm z}$ and $\matr A^\dagger \matr A$ converge almost surely to some limiting spectra whose supports are compact. 
\end{assumption}
Due to lack of an explicit definition of the ``Lagrangian" matrix $\matr \Lambda$, Assumption~2 is rather implicit. Nevertheless it can be considered in the same vein as the so-called thermodynamic limit in statistical physics: all microscopic variables converge to deterministic values in the thermodynamic limit \cite{Mezard}.

For example, under Assumption~1-a) and Assumption~2, it turns out that $\matr\Lambda_{\rm x}$ and $\matr J_{\rm z}\triangleq \matr A^\dagger\matr \Lambda_{\rm z}\matr A$ are asymptotically free \cite{Collins} and from \cite[Lemma~ 3.3.4]{hiai} we have that\footnote{In fact we can define the R-transform on negative real line. However in the exposition it requires an implicit assumption that $\matr \Lambda$ is being positive-definite.}
\begin{equation}
{\rm R}^K_{\matr \Lambda_{\rm x}+\matr J_{\rm z}}(\omega) \simeq {\rm R}^K_{\matr \Lambda_{\rm x}}(\omega)+{\rm R}_{\matr J_{\rm z}}^K(\omega), \quad \Im \omega<0.  \label{key}
\end{equation}
Here for a $T\times T$ symmetric matrix $\matr X$ ${\rm R}^T_{\matr X}$ denotes the R-transform of the spectrum of $\matr X$ (see Appendix~\ref{pre}) and 
${\simeq}$ stands for the large system approximation that turns to an almost surely equality in the large system limit. Furthermore we introduce 
\begin{equation}
{\mathcal R}^{T}_{\matr X}(r)\triangleq\lim_{\omega\to r}\Re{\rm R}^{T}_{\matr X}(\omega), \quad \Im r=0
\end{equation} 
whenever the limit exists.

It turns out that by \emph {solely} invoking ``additive free convolution", e.g. \eqref{key}, we can easily solve the complexity issue of the fixed point identities for $\matr \LL_{\rm x}$ and $\matr \LL_{\rm z}$ which do not require matrix inversion. First we consider the simplification for $\matr \LL_{\rm x}$. To than end let us first define the auxiliary variable 
\begin{align}
q\triangleq\frac{1}{K} {\rm tr}\{(\matr \Lambda_{\rm x}+\matr J_{\rm z})^{-1}\}=\frac{1}{K}\sum_{k\in \mathcal K}\frac{1}{[\Lambda_{\rm x}]_{kk}+[\LL_{\rm x}]_{kk}} \label{defq}.
\end{align}
Then by invoking \eqref{key} we easily obtain that (see Appendix~\ref{Apadf}) 
\begin{equation}
q \simeq \frac{1}{K}\sum_{k\in \mathcal K} \frac{1}{[\Lambda_{\rm x}]_{kk}+{\mathcal R}^{K}_{\matr J_{\rm z}}(-q)}. \label{adf}
\end{equation}
Thereby, we conclude that
\begin{align}
[\LL_{\rm x}]_{kk}\simeq{\mathcal R}^K_{\matr J_{\rm z}}(-q), \quad  k\in \mathcal K.\label{Self1}
\end{align}
The average of \eqref{Self1} over the random matrix $\matr A$ agrees with \cite[Eq.~(50)]{Ole}. Note that the simplification in \eqref{Self1} is still implicit due to the definition of $q$ in \eqref{adf}. Subsequently we present an explicit complexity simplification for $[\LL_{\rm x}]_{kk}$. First we note that \eqref{Self1} states that we can replace all the elements $[\LL_{\rm x}]_{kk}$, $k\in \mathcal K$ by a single scalar quantity, say $\LL_{\rm x}$. This allows us to write $q\simeq\left<\matr \sigma_{\rm x}(\matr \kappa_{\rm x}, \mathcal \LL_{\rm x}\matr {\bf I})\right>$ with $\matr\kappa_{\rm x}=\mathcal \LL_{\rm x}^{-1}\matr A^\dagger\matr m+\matr{\hat x}$. Then, from \eqref{Self1} we write an explicit fixed point identity for $\LL_{\rm x}$ as
\begin{equation}
\LL_{\rm x}={\mathcal R}^K_{\matr J_{\rm z}}(-\left<\matr \sigma_{\rm x}(\matr \kappa_{\rm x}; \mathcal \LL_{\rm x}\matr {\bf I})\right>).\label{S1}
\end{equation}

As a second part we address similar complexity simplification for $[\LL_{\rm z}]_{nn}$ for $n \in \mathcal N$. To that end let us introduce an auxiliary $N\times 1$ vector $\tilde{\matr\tau}_{\rm m}$ whose entries are defined as
\begin{align}
[\tilde\tau_{\rm m}]_{n}&\triangleq[\Lambda_{\rm z}]_{nn}-[\Lambda_{\rm z}]^2_{nn}[A(\Lambda_{\rm x}+A^\dagger\Lambda_{\rm z} A)^{-1}A^\dagger]_{nn}\label{lambdan}\\
&=[(\Lambda_{\rm z}^{-1}+A\Lambda_{\rm x}^{-1} A^\dagger)^{-1}]_{nn},\label{mi}
\end{align}
where \eqref{mi} follows directly from Woodbury's matrix inversion lemma. Furthermore by making use of \eqref{f2} for \eqref{lambdan} we can write the following fixed-point identity
\begin{align}
[(\Lambda_{\rm z}^{-1}+A\Lambda_{\rm x}^{-1} A^\dagger)^{-1}]_{nn}&=[\Lambda_{\rm z}]_{nn}-\frac{[\Lambda_{\rm z}]^2_{nn}}{[\Lambda_{\rm z}]_{nn}+[\LL_{\rm z}]_{nn}} \label{good1}\\
&= \frac{1}{[\Lambda_{\rm z}^{-1}]_{nn}+[\LL_{\rm z}^{-1}]_{nn}}. \label{good2}
\end{align}
Thus, we can invoke identical arguments on the additive free convolution approximation above for $[\LL_{\rm z}]_{nn}$ as well. Specifically, under Assumption~1-b) and Assumption~2, for a large $N,K$ we have 
\begin{equation}
[\LL_{\rm z}]_{nn}\simeq\frac{1}{{\mathcal R}^N_{\matr J_{\rm x}}(-\left<\tilde{\matr\tau}_{\rm m}\right>)}, \quad  n \in \mathcal N\label{Self2}
\end{equation} 
with $\matr J_{\rm x}\triangleq \matr A\matr\Lambda_{\rm x}^{-1} \matr A^\dagger$.  The complexity simplification \eqref{Self2} is still implicit due the definition of $\tilde {\matr \tau}_{\rm m}$. To present an explicit form of it consider first \eqref{good1} and \eqref{good2} such that we can write 
\begin{align}
[\tilde{\tau}_{\rm m}]_{n}&= [\LL_{\rm z}]_{nn}-\frac{[\LL_{\rm z}]^2_{nn}}{[\LL_{\rm z}]_{nn}+[\LL_{\rm z}]_{nn}}\\
&= [\LL_{\rm z}]_{nn}\left(1-[\LL_{\rm z}]_{nn}[\sigma_{\rm z}(\matr \kappa_{\rm z}; \matr \LL_{\rm z})]_{n} \right).
\end{align}
On the other hand, \eqref{Self2} implies that we can replace all the elements $[\LL_{\rm z}]_{nn}$, $n\in \mathcal N$ by a single scalar quantity, say $\LL_{\rm z}$. Now for convenience let us define $N\times 1$ vector $\matr\tau_{\rm m}$ whose entries are given by 
\begin{equation}
[{\tau}_{\rm m}]_{n}\triangleq\LL_{\rm z}\left(1-\LL_{\rm z}[\sigma_{\rm z}(\matr \kappa_{\rm z};\LL_{\rm z} \matr {\bf I})]_{n} \right), \quad n \in \mathcal N.
\end{equation}
Then following \eqref{Self2} we introduce an explicit fixed-point identity for $\LL_{\rm z}$ as 
\begin{equation}
\LL_{\rm z}=\frac{1}{{\mathcal R}^N_{\matr J_{\rm x}}(-\left<\matr{\tau}_{\rm m}\right>)}. \label{S2}
\end{equation}

So far we have shown in \eqref{S1} and \eqref{S2} how to bypass the need for matrix inversion to ``update" $\matr \Lambda_{\rm x}$ and $\matr \Lambda_{\rm z}$, respectively. However this treatment require solving a highly non-trivial random matrix problem i.e. deriving the closed form solution for ${\mathcal R}^K_{\matr J_{\rm z}}$ and ${\mathcal R}^N_{\matr J_{\rm x}}$. This is usually, though not always, not possible. On the other hand deriving the solution of e.g ${\mathcal R}^K_{\matr J_{\rm z}}$ in the limiting case, denote ${\mathcal R}_{\matr J_{\rm z}}$, is rather simpler. Due to the uniform convergence property of the R-transform \cite[Lemma~ 3.3.4]{hiai}, this approach would allow us to accurately predict, for example ${\mathcal R}^K_{\matr J_{\rm z}}$, for large $N,K$. This is what we show in the next subsection for the zero mean iid Gaussian matrix ensemble.
\subsubsection*{Example: The zero-mean and iid case, i.e. GAMP}
In this section we provide the explicit solutions for $\LL_{\rm x}$ and $\LL_{\rm z}$ when the entries of $\matr A$ are assumed to be iid Gaussian with zero mean and variance $1/N$. 

From the well-known Mar\u {c}henko-Pastur theorem we obtain that (see Appendix~\ref{Apiid})
\begin{align}
{\rm R}^K_{\matr J_{\rm z}}(\omega)&\simeq \frac{1}{N}\sum_{n\in \mathcal N} \frac{1}{[\Lambda_{\rm z}^{-1}]_{nn}-\omega/\alpha}\label{iida}  \\
{\rm R}^N_{\matr J_{\rm x}}(\omega)&\simeq \frac{1}{\alpha K}\sum_{k\in \mathcal K} \frac{1}{[\Lambda_{\rm x}]_{kk}-\omega}.\label{iidb}
\end{align}
Then we obtain the following expression for $\LL_{\rm x}$ and $\LL_{\rm z}$ as
\begin{align}
\LL_{\rm x}&\simeq \frac{1}{N}\sum_{n\in \mathcal N} \frac{1}{[\Lambda_{\rm z}^{-1}]_{nn}+\left<\matr \sigma_{\rm x}(\matr \kappa_{\rm x}; \mathcal \LL_{\rm x} \matr {\bf I})\right>/\alpha} \\
\frac{1}{\LL_{\rm z}}&\simeq \frac{1}{\alpha K}\sum_{k\in \mathcal K} \frac{1}{[\Lambda_{\rm x}]_{kk}+\left<\matr \tau_{\rm m}\right>}. 
\end{align}
From these equations one can conclude that 
\begin{align}
\LL_{\rm z}\simeq \frac{\alpha}{\left<\matr \sigma_{\rm x}(\matr \kappa_{\rm x}; \mathcal \LL_{\rm x}\matr {\bf I})\right>}, \quad \LL_{\rm x}\simeq {\left<\matr\tau_{\rm m}\right>}.
\end{align}
Thus we recover the fixed point of the GAMP-2nd order updates for the zero-mean iid matrix ensemble as in \eqref{iid2}.

\section{Conclusion}
For the given zero-mean iid Gaussian matrix ensemble, the fixed points of GAMP ``asymptotically'' coincide with the stationary points of Gibbs free energy under first- and second-moment constraints. It turns out that the only critical issue for GAMP is the update rules for "variance" parameters $\matr \LL_{\rm x}$ and $\matr \LL_{\rm z}$. These parameters play a central role. Specifically a crude update rule for a given measurement matrix ensemble would completely spoil the optimality of the algorithm. 
If for general invariant matrix ensembles, $\matr \LL_{\rm x}$ and $\matr \LL_{\rm z}$ can be updated based on the R-transform formulation in \eqref{S1} and \eqref{S2}; the algorithm ``asymptotically'' fulfills the stationary points identities of Gibbs free energy formulation. Once the closed form expressions of \eqref{S1} and \eqref{S2} are obtained, the resulting algorithm includes solely $O(N)$ operations. But the computation of the solutions to these identities is not trivial. Nevertheless it is sometimes doable, e.g. the random row orthogonal matrix ensembles. Furthermore once either the prior or the likelihood is expressed in terms of a Gaussian function, the R-transform formulation becomes rather trivial. In general updating $\matr \LL_{\rm x}$ and $\matr \LL_{\rm z}$ requires a matrix inversion at each iteration, e.g. see \eqref{EP1}--\eqref{EP8}. An alternative, but sub-optimal, method would be the Swept-AMP algorithm \cite{swept} that is based on the GAMP methodology and includes $O(N^2)$ operations.

\bibliographystyle{IEEEtran}
\bibliography{IEEEabrv,liter}

\appendices
\section{Preliminaries}\label{pre}
Let $\rm {P}_{\rm X}$ a probability distribution on real line. We denote the Stieltjest transform of ${\rm P}_X$ as  
\begin{equation}
{\rm G}_X(s)\triangleq \int \frac{{\rm dP}_X(x)}{x-s}, \quad \Im s>0
\end{equation}
where $\Im {\rm G}_X(s)>0$ \cite{Hackem}.

The R-transform of ${\rm P}_{X}$ is defined as \cite{hiai}
\begin{equation}
{\rm R}_{X}(\omega)\triangleq {\rm G}^{-1}_{X}(-\omega)-\frac{1}{\omega}, \quad \Im \omega <0 \label{Rdef}
\end{equation}
with ${\rm G}^{-1}_{X}$ denoting the inverse of ${\rm G}_{X}$. Equivalently,
\begin{equation}
{\rm R}_{X} (-{\rm G}_{X} (s))=s+\frac{1}{{\rm G}_{X}(s)}.\label{Rtrans}
\end{equation}
Here we draw the attention of the reader that $\Im{\rm R}_{X}(\omega)< 0$, for $\Im \omega<0$ unless ${\rm P}_{X}$ is a Dirac distribution. This fact follows from the following property of the Stieltjest transform \cite[Proposition~2.2]{Hackem}: for $\Im s>0$, $\Im\{\frac{1}{{\rm G}_X(s)}+s\}\leq 0$ where the equality holds if, and only if, ${\rm P}_{X}$ is a Dirac distribution.
\begin{remark}
Let ${\rm P}_{X}$ have a pdf ${\rm p}_{\rm X}$.
Furthermore let ${\rm p}_{X}(0)=0$; so that  $\lim_{\epsilon\to 0^+}\Im{\rm G}_{\rm X}(j\epsilon)=0$. Moreover let 
\begin{equation}
q\triangleq\lim_{\epsilon\to 0^{+}}{\rm G}_{\rm X}(j\epsilon)=\int x^{-1}{\rm dP}_{X}(x)<\infty. \label{inversemean}
\end{equation}
Then we have the following identity
\begin{equation}
\frac{1}{q}=\lim_{\omega\to q} {\rm R}_{X}(-\omega).
\end{equation}
\end{remark}
Consider an $T\times T$ symmetric matrix $\matr X$. Let $\mathcal L$ be the set containing the eigenvalues of $\matr X$. The spectrum of $\matr X$ is denoted by 
\begin{equation}
{\rm P}_{\matr X}^{T}(x)\triangleq \frac{1}{T}\left\vert \{\lambda \in \mathcal L: \lambda<x \}\right\vert. 
\end{equation}
We denote the Stieltjest transform and the R-transform of ${\rm P}_{\matr X}^T$ by ${\rm G}_{\matr X}^T$ and ${\rm R}_{\matr X}^T$, respectively. Furthermore if for $T\to \infty$, $\matr X$ has a limiting spectrum almost surely it is denoted by ${\rm P}_{\matr X}$. Moreover, the Stieltjest transform and the R-transform of ${\rm P}_{\matr X}$ are denoted by ${\rm G}_{\matr X}$ and ${\rm R}_{\matr X}$, respectively.
\section{Proof of \eqref{adf}}\label{Apadf}
Note that from definition in \eqref{defq} we have 
\begin{align}
q=\int x^{-1}{\rm dP}_{\matr \Lambda_{\rm x}+ \matr J_{\rm z}}^K(x) \label{key0}.
\end{align}
By invoking Remark~1 and \eqref{key} (under the Assumption~1-a) and Assumption~2), successively we can write
\begin{align}
&\lim_{\epsilon \to 0^+}{\rm R}^{K}_{\matr \Lambda_{\rm x}}(-q_{\epsilon}-j\epsilon)+{\rm R}^{K}_{\matr J_{\rm z}}(-q_{\epsilon}-j\epsilon)-\frac{1}{q_{\epsilon}+j\epsilon}\simeq 0. \label{key2}
\end{align}
Here, without loss of generality, we can define $q_{\epsilon}\triangleq q+\epsilon$. On the other hand, from the definition of the R-transform in \eqref{Rtrans} we have 
\begin{equation}
{\rm R}^K_{\matr \Lambda_{\rm x}}(-{\rm G}^K_{\matr \Lambda_{\rm x}}(-s))+ s- \frac{1}{{\rm G}^K_{\matr \Lambda_{\rm x}}(-s)}=0 \quad \Im s<0.
\end{equation}
Hence we can write
\begin{align}
q_{\epsilon}+j\epsilon&\simeq {\rm G}^K_{\matr \Lambda_{\rm x}}(-{\rm R}^{K}_{\matr J_{\rm z}}(-q_\epsilon-j\epsilon)) \\
 &= \frac{1}{K}\sum_{k\in \mathcal K} \frac{1}{[\Lambda_{\rm x}]_{kk}+{\rm R}^{K}_{\matr J_{\rm z}}(-q_\epsilon-j\epsilon)}. \label{key3}
\end{align}
This completes the proof.

\section{Proof of \eqref{iida} \& \eqref{iidb}} \label{Apiid}
Let us first consider $\matr J_{\rm z}=\matr A^\dagger \matr \Lambda_{\rm z} \matr A$. Note that we do not assume that $\matr A$ and $\matr \Lambda_{\rm z}$ are independent. On the other hand, Assumption~2 results in that $\matr A^\dagger \matr A$ and $\matr \Lambda_{\rm z}$ are asymptotically free of each others. In this way we can find ${\rm R}_{\matr J_{\rm z}}$ by means of the so-called multiplicative free convolution \cite{Rao}. However this requires the reader to be familiar with the S-transform in free probability. In fact, by invoking standard random matrix results we can bypass the need for using the S-transform. Specifically, from the well-known Mar\u {c}henko-Pastur theorem, we can write
\begin{equation}
{\rm G}_{\matr J_{\rm z}}(s)=\frac{1}{-s+\int \frac{{\rm dP}_{\matr \Lambda_{z}}(x)}{1/x+{\rm G}_{\matr J_{\rm z}}(s)/\alpha}}. \label{Marchenko}
\end{equation}
The result \eqref{Marchenko} is proven under the assumptions that the entries of $\matr A$ are iid (not necessarily Gaussian) with zero mean and $\matr A$ is independent of $\matr \Lambda_{\rm z}$ \cite{Silverstein}. Due to the asymptotic freeness, this result holds when the entries are restricted to Gaussian but without restriction that $\matr A$ and $\matr \Lambda_{\rm z}$ are independent. Now by letting $s={\rm G}_{\matr J_{\rm z}}^{-1}(-\omega)$ in \eqref{Marchenko} and from the definition of the R-transform in \eqref{Rdef} we have
\begin{equation}
{\rm R}_{\matr J_{\rm z}}(\omega)=\int \frac{{\rm dP}_{\matr \Lambda_{\rm z}}(x)}{1/x-\omega/\alpha}. \label{F1}
\end{equation}
Furthermore following the identical arguments for $\matr J_{\rm x}$ we find
\begin{equation}
{\rm R}_{\matr J_{\rm x}}(\omega)=\frac{1}{\alpha}\int \frac{{\rm dP}_{\matr \Lambda_{\rm x}}(x)}{x-\omega}. \label{F2}
\end{equation}
Due to \cite[Lemma~ 3.3.4]{hiai}, the right hand side of the expressions in \eqref{iida} and \eqref{iidb} converge uniformly to \eqref{F1} and \eqref{F2}, respectively. This completes the proof.
\end{document}